\documentclass[a4paper,11pt]{article}
\usepackage{amsmath,amssymb,mathtools,cite,graphicx,fancyhdr}
\usepackage{bbold,color,mathpazo,authblk}
\usepackage{caption}
\graphicspath{{Figures/}}
\newcommand{\dd}{\text{d}}
\newcommand{\ii}{\text{i}}
\newcommand{\ee}{\text{e}}

\usepackage{hyperref}
\hypersetup{
	unicode=false,          % non-Latin characters in Acrobat’s bookmarks
	pdftitle={Resonant Hawking radiation as an instability},    % title
	pdfauthor={David Bermudez},     % author
	pdfsubject={Analogue Gravity},   % subject of the document
	pdfcreator={David Bermudez},   % creator of the document
	pdfproducer={David Bermudez}, % producer of the document
	pdfkeywords={theory of instabilities, analogue gravity, Hawking radiation, Bogoliubov-De Gennes equation, black hole laser, quantum, physics, mathematics}, % list of keywords
	pdfnewwindow=true,      % links in new window
	colorlinks=true,       	% false: boxed links; true: colored links
	linkcolor=magenta,      % color of internal links
	citecolor=blue, 			% color of links to bibliography
	filecolor=red,      		% color of file links
	urlcolor=red           	% color of external links
}

\begin{document}
\lhead{Resonant Hawking radiation as an instability}
\rhead{Bermudez, Leonhardt}
%\cfoot{\page}

\title{Resonant Hawking radiation as an instability}
\author{David Bermudez\footnote{{\it email:} dbermudez@fis.cinvestav.mx}}
\affil{\textit{Departamento de F\'{\i}sica, Cinvestav, A.P. 14-740, 07000 Ciudad de M\'exico, Mexico}}
\author{Ulf Leonhardt\footnote{{\it email:} ulf.leonhardt@weizmann.ac.il}}
\affil{\textit{Department of Physics of Complex Systems, Weizmann Institute of Science.}\\ \textit{761001 Rehovot, Israel}}

\renewcommand\Authands{, and }

\date{}
\maketitle

\begin{abstract}
We consider a simple model for a black-hole laser: a Bose-Einstein condensate with uniform speed of sound and partially uniform flow, establishing two horizons, a black-hole and a white-hole horizon. Waves confined between the horizons are amplified similar to radiation in a laser cavity. Black-hole lasing appears as an instability with discrete sets of modes given approximately by a round-trip condition. We found that, in addition to the regular Hawking radiation, trans-Planckian radiation does tunnel out of the black-hole laser.\\

\noindent{\it Keywords}: theory of instabilities; analogue gravity; Hawking radiation; Bogoliubov-De Gennes equation; black hole laser;
\end{abstract}

\section{Introduction}
Do black holes radiate? \cite{Helfer2003} Even though there has been intensive research to answer this question in the last 44 years, Hawking radiation \cite{Hawking1974} remains still undetected. However, this has not stopped scientists from trying to understand this phenomenon. In particular, Unruh \cite{Unruh1981} proposed an effect analogous to a black hole emitting light (photons) with a fluid emitting sound waves (phonons). The study of analogue Hawking radiation and other effects usually related to gravity has given birth to a new and intensive area of research: analogue gravity. With this approach, new ways of thinking have opened both in gravity and in the analogue theories, e.g., in condensed matter physics\cite{Garay2000,Barcelo2011}, hydrodynamics \cite{Rousseaux2008,Euve2016}, optics \cite{Philbin2008,Bermudez2016var}, among others.

In this paper, we consider a trans-sonic Bose-Einstein condensate (BEC) as our quantum fluid, as it is a system described by a simple theory while still offering good perspectives for demonstrating the quantum effects of event horizons in the laboratory. Furthermore, we study an arrangement known as ``black-hole laser'' (BHL) \cite{Corley1999}, which consists of a finite stationary transonic medium, i.e., a medium with a spatially nonuniform flow that varies from subsonic to supersonic and back to subsonic speeds, establishing two horizons: the white-hole (WH) and black-hole (BH) horizons. These horizons confine Hawking radiation in the space between them and, as they create the Hawking radiation in the first place, they act as both the cavity mirrors and the gain medium of a laser: the black-hole laser. 

There has been substantial literature on the theory of black-hole lasing \cite{Coutant2010,Finazzi2010njp,Faccio2012,Finazzi2015prl,Peloquin2016,GaonaReyes2017} and a disputed \cite{Leonhardt2018} experimental demonstration \cite{Steinhauer2014}. Here we develop the simplest possible model of a BHL, which allows us to draw a conclusion that is normally hidden in the technicalities of more complicated models closer to experimental details. We are going to show that not only regular Hawking radiation is emitted by the BHL, but also trans-Planckian radiation. This is radiation characterized by wavelengths comparable with the Planck scale of the problem. In black-hole analogues, this scale is set by dispersion \cite{Jacobson1991}. In the case of BECs, it is given by the healing length of the condensate. Normally, this trans-Planckian radiation cannot propagate beyond the black-hole horizon nor leave the white-hole, but in the case of instabilities, as in the BHL, it can.

An instability is characterized by a complex frequency $\omega$ with positive imaginary part\footnote{We adopt the convention of physics that stationary waves oscillate with $\exp(-i\omega t)$, whereas in engineering they tend to oscillate with $\exp(j\omega t)$.}. In this case, propagating plane waves must also have a complex wavenumber $k$. What should be the sign of the imaginary part of $k$? Imagine the source of the instability --- in our case the BHL --- emits a wave growing in time. Consider two observers at equal time, one closer to the source and the other farther away. As the radiation takes time to propagate, the closer observer must perceive a larger amplitude than the more distant one. Therefore, the spatial part $\exp(i k x)$ of the emitted wave must decay while propagating, which implies that the imaginary part of $k$ must be positive for positive $x$. For negative $x$ (on the other side of the source), the imaginary part of $k$ must be negative. 

The waves made by the instability are thus exponentially confined in space while exponentially growing in time (until the resources are getting depleted). Mathematically, they form confined, square-integrable modes similar to the eigenfunctions of a potential well. Physically, however, they are not confined at all. Quite the opposite, they describe the growing flux of radiation propagating away from the instability. The same concept also applies to waves that, for real $\omega$, are not propagating, but are exponentially confined. For complex $\omega$ with positive imaginary part, they behave in the same way as the former propagating waves. For real frequencies they are evanescent waves, for complex frequencies they are propagating, as the distinction between propagating and evanescent waves ceases to exist in this case. In our case, this implies that trans-Planckian radiation can tunnel out of the confining horizons and escape. As it turns out for our model, the trans-Planckian component is comparable in magnitude to the regular Planckian one. 

We also found that in our model a substantial part of Hawking radiation is reflected at the horizons, i.e., turned from counter-propagating to co-propagating.\footnote{Reflection is frequently called $u$-$v$ mixing in the literature of black-hole analogues.} This is not surprising, as we consider a model with sharp interfaces where reflections are prevalent. We are also going to show that the modes of the BHL are well described by a simple, intuitive model typical for radiation in cavities. The phase of a round-trip inside the cavity should be a multiple of $2\pi$ plus two times $\pi/2$ from the two turning points, the horizons. Requiring thus that the phase of a round-trip for real frequencies is an odd multiple of $\pi$ turns out to give an excellent approximation for the real part of the wavenumber, despite amplification, tunneling and reflection involved in black-hole lasing. 

\section{Equations of motion}

\subsection{Bogoliubov dispersion}

The theory of elementary excitations in BECs \cite{Pitaevskii2016} relies on linearizing the quantum field of atoms around the classical mean field, which gives the Bogoliubov-de Gennes equations \cite{Pitaevskii2016} for the two components of the elementary excitations. For instabilities, this theory was developed in Ref.~\cite{Leonhardt2003}. Here we perform a simplification: the only ingredient we need from the Bogoliubov-de Gennes equations is the Bogoliubov spectrum \cite{Pitaevskii2016}:
\begin{equation}
	\omega^2=c^2k^2+\frac{\hbar^2}{4m^2}k^4
\end{equation}
where $m$ is the atomic mass of the condensate and $c$ the speed of sound, which can be rewritten in terms of the single constant $k_0$ as
\begin{equation}
	\omega^2=c^2k^2\left(1+\frac{k^2}{k_0^2}\right)=F^2(k).
	\label{bogo}
\end{equation}
The constant $k_0$ quantifies the deviation of the dispersion relation from the ideal dispersion of a linear wave (from relativistic wave propagation with $c$ playing the role of the speed of light). It is related to the so-called healing length $\xi$ \cite{Pitaevskii2016} by $\xi^2=2/k_0^2$. The healing length $\xi$ defines the effective Planck scale for analogues of gravity in BECs. 

\subsection{Quantum field}

Unruh proved \cite{Unruh1981} that the equation for a small perturbation $\phi$ of the velocity potential --- a sound wave --- in a rotation-less fluid with a given velocity profile $v(x)$ is 
\begin{equation}
	(\partial_t+\partial_x v)(\partial_t +v\partial_x)\phi =\partial_x^2 \phi.
\end{equation}
We can generalize this equation by including the Bogoliubov dispersion given in Eq. \eqref{bogo}, and get
\begin{equation}
	(\partial_t+\partial_x v)(\partial_t +v\partial_x)\phi =\partial_x^2 \phi-\frac{1}{k_0^2}\partial_x^4\phi.
\end{equation}
Alternatively, this equation of motion can be obtained by proposing an appropriate action for the field \cite{Corley1999,GaonaReyes2016}, which in this case takes the form
\begin{equation}
	S_\phi = \frac{1}{2}\int \dd^2x\left([(\partial_t+v\partial_x)\phi]^2+\phi\left[\partial_x^2 \phi-\frac{1}{k_0^2}\partial_x^4\right]\phi\right).
\end{equation}
The operator of the right-hand side is the sum of two terms. The first one comes from the ordinary relativistic action, and the second one includes a higher derivative that produces superluminal group velocities due to its sign. The corresponding dispersion relation is then
\begin{equation}
	(\omega - v k)^2=F^2(k)=c^2k^2\left(1+\frac{k^2}{k_0^2}\right),
	\label{disp}
\end{equation}
where derivatives of both the velocity profile and the wave number were neglected.

\begin{figure}
	\centering
	\includegraphics[scale=0.5]{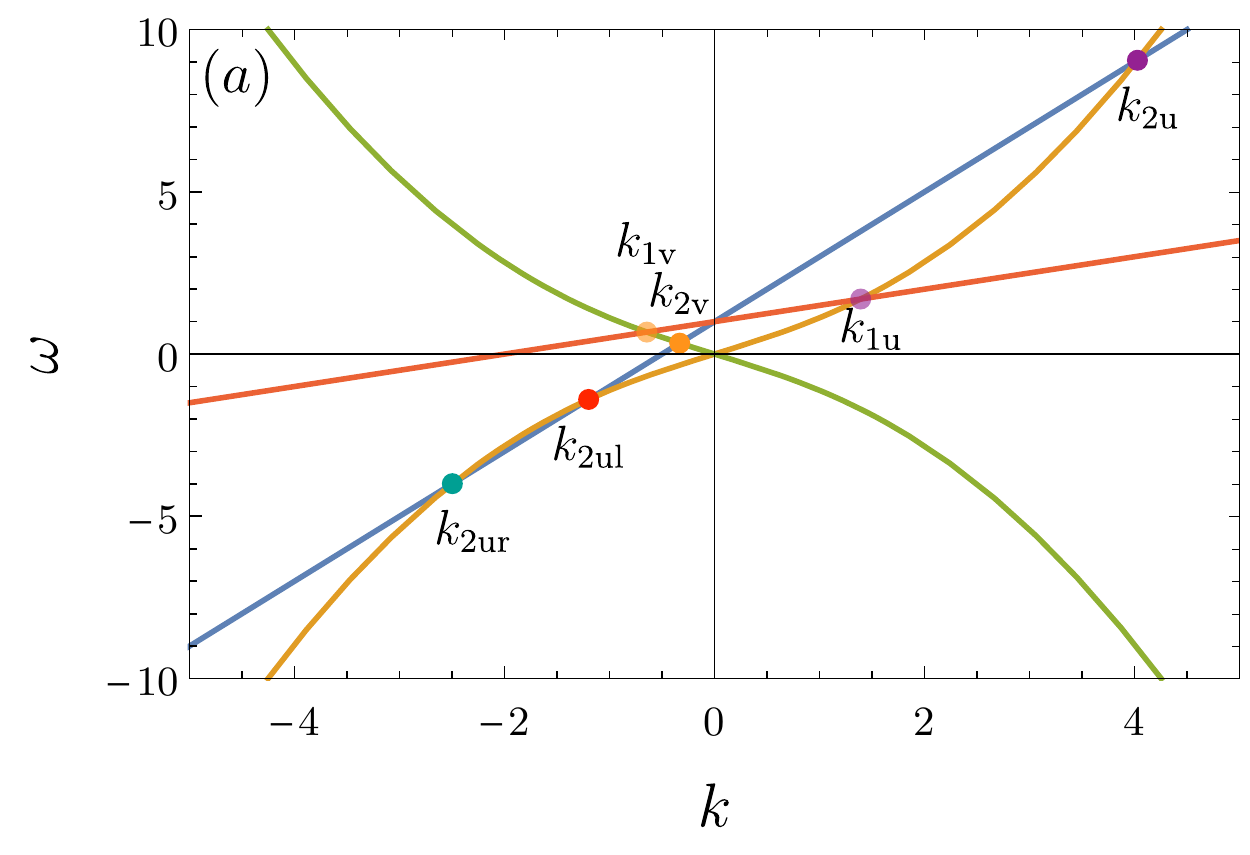}\hspace{5mm}
	\includegraphics[scale=0.26]{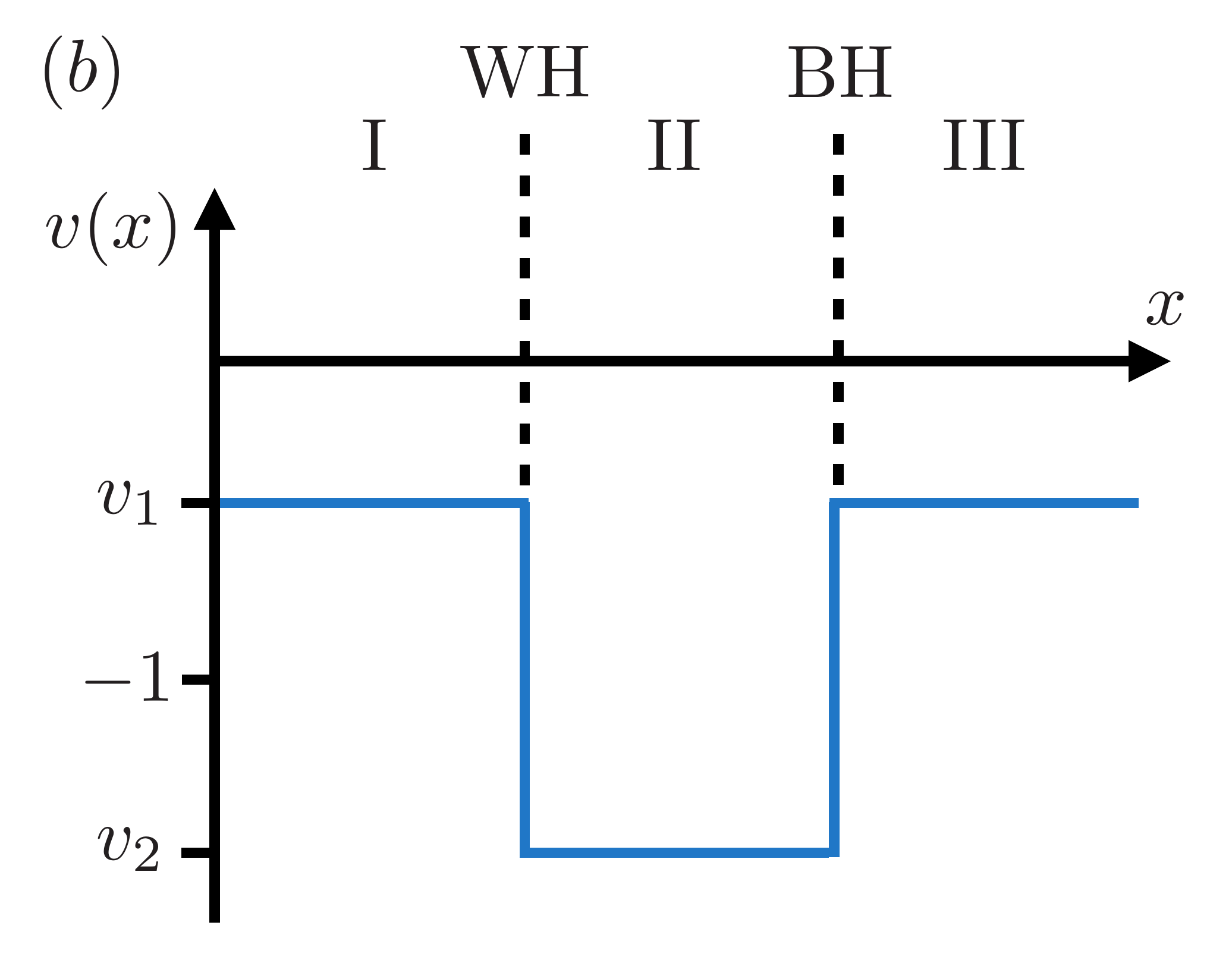}
	\caption{\small{(a) Dispersion relation for counter-propagating (orange) and co-propagating waves (green). Graphical solution for two velocity profiles: one subsonic $v_1$ (red) and one supersonic $v_2$ (blue). Here $\omega=1$, $k_0=2$, $c=1$, $v_1=-1/2$, and $v_2=-2$. (b) Fluid velocity with three distinct regions, I and III with subsonic velocity $-1<v_1<0$ and II with supersonic velocity $v_2<-1$.}}
	\label{figdisp}
\end{figure}

This dispersion relation is visualized in Fig. \ref{figdisp} (a) where we plot the square root of each side of Eq. \eqref{disp} separately. We choose the fluid flow to the left, i.e., $v<0$. The positive root of the right-hand side (orange curve) gives the solutions moving to the right or counter-propagating with the flow, and the negative root (green curve) gives the solutions moving to the left or co-propagating with the flow. The left-hand side are straight lines with two velocities, $v_1$ (subsonic) and $v_2$ (supersonic).

\section{Black hole laser with real frequencies}

For a given $\omega$, we can find up to four solutions for the dispersion relation in Eq. \eqref{disp} through a numerical (or graphical) method and each velocity. We also show in Fig. \ref{figdisp} the solutions in the dispersion relation diagram for two velocities: for a supersonic velocity $v_2$ we find four solutions (labeled with a subscript 2) and for a subsonic velocity $v_1$ only two (subscript 1).

Moreover, the solutions are also labeled with subscripts ``u'' for a counter-propagating mode moving to the right with $k>0$ that exists for both $v_1$ and $v_2$, ``ur'' and ``ul'' for counter-propagating solutions with $k<0$ moving to the left and to the right, respectively, and ``v'' for co-propagating solutions also with $k<0$ moving to the left.

\begin{figure}
	\centering
	\includegraphics[scale=0.27]{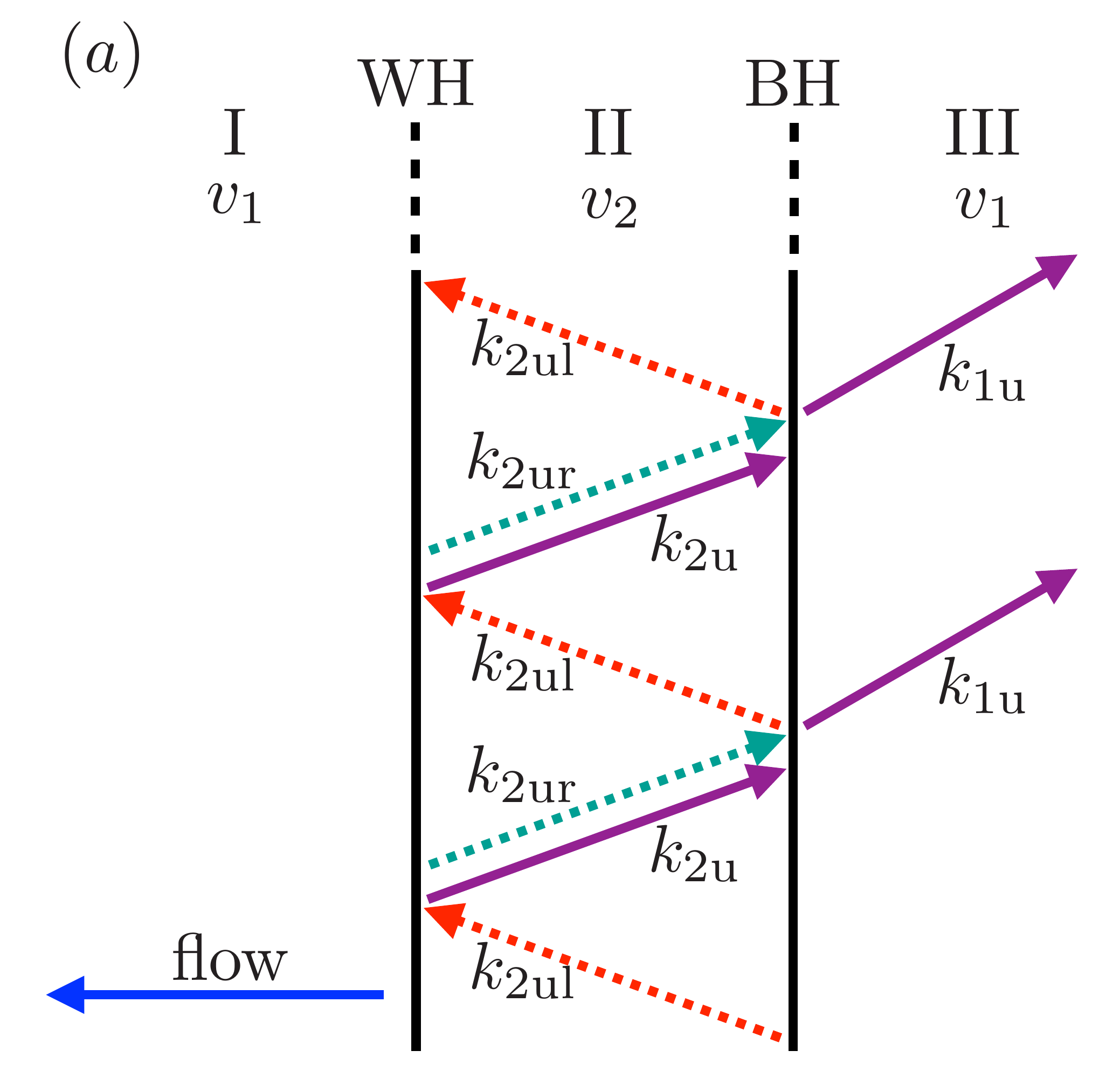}\hspace{5mm}
	\includegraphics[scale=0.27]{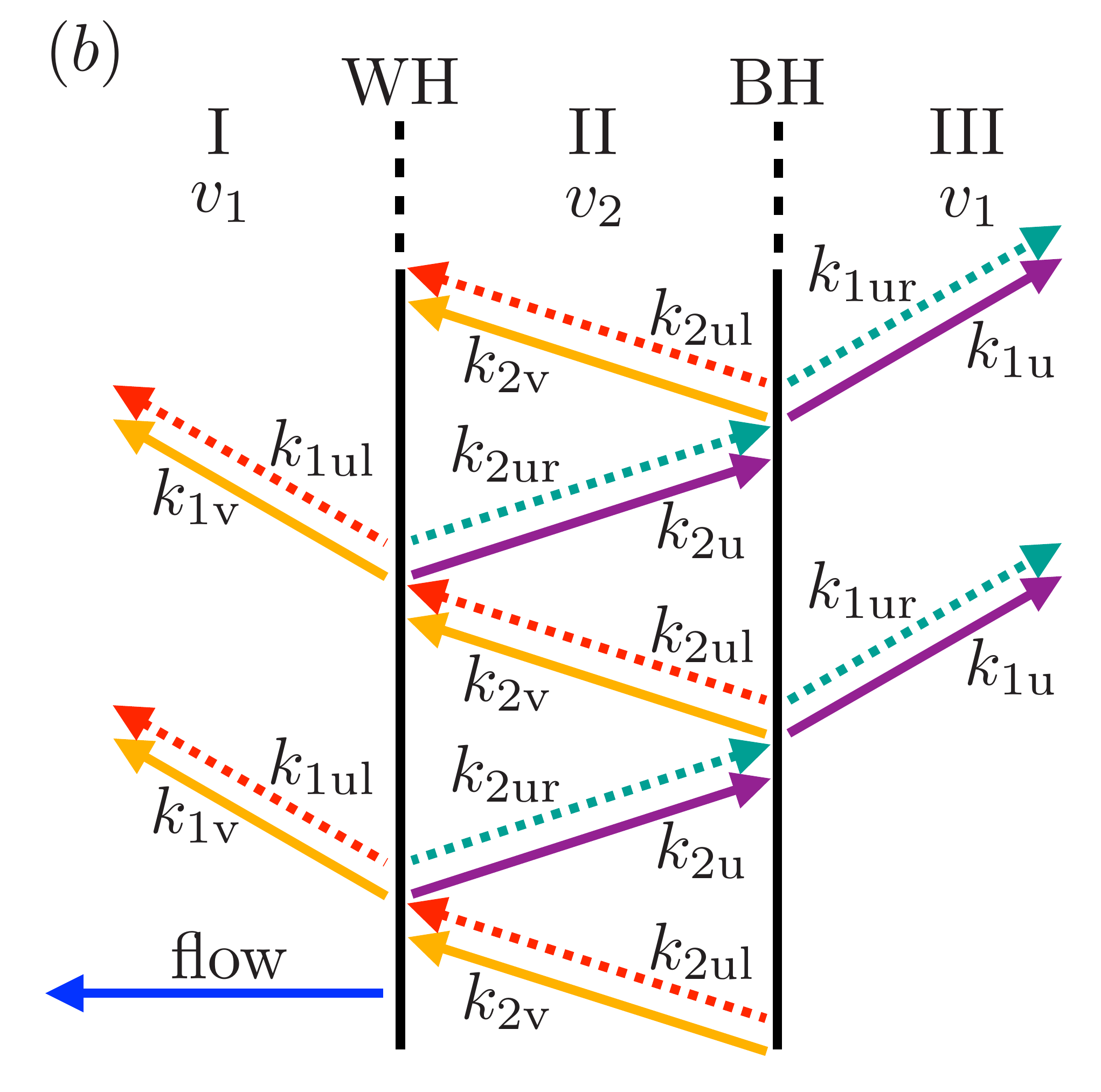}
	\caption{\small{(a) Diagram of the evolution of trapped mode $k_\text{2ul}$ for real frequencies. The norm of the outgoing mode $k_\text{1u}$ increases in each cycle by the lasing effect of the system. (b) For complex frequencies, the evolution includes other outgoing modes that are exponentially decaying away from the cavity.}}
	\label{figbhl}
\end{figure}

The black-hole laser is a system where the flow velocity changes from a subsonic velocity $v_1$ (region I) to a supersonic one $v_2$ (region II) only for a finite distance $L$, to then return to $v_1$ (region III), such velocity profile can be seen in Fig. \ref{figdisp} (b). The finite supersonic region is the cavity. We are considering a simple system where the flow velocity is constant everywhere, i.e., step-like or flat-profile configuration \cite{Larre2012}. If $k_0\neq 0$, for any $\omega$ there is a domain of $v_1$ and $v_2$ where the solutions have the behavior shown in Fig. \ref{figdisp}(a), and modes $k_\text{2ul}$ and $k_\text{2ur}$ are trapped inside the cavity by the difference in fluid velocity, creating an analogue to the Hawking effect (see Fig. \ref{figbhl}(a)). The left side of the cavity is considered an analogue of a WH and the right side an analogue of a BH.

\subsection{Direction of travel}
\begin{figure}
	\centering
	\includegraphics[scale=0.6]{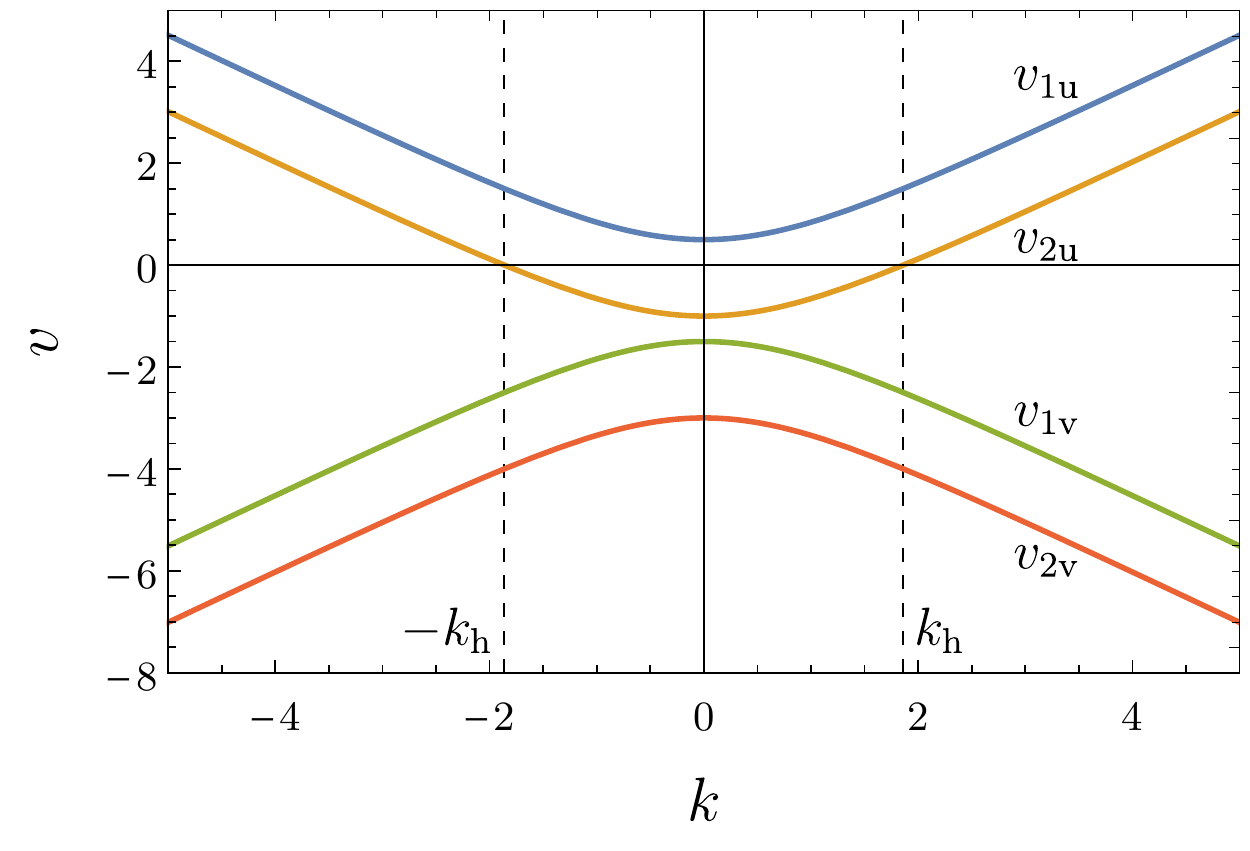}
	\caption{\small{Group velocity of modes $k$ after considering fluid flow, given by Eq. \eqref{vels} for $v_1=-1/2$ and $v_2=-2$. From highest to lowest they are: $v_\text{1u}$ (blue), $v_\text{2u}$ (orange), $v_\text{1v}$ (green), $v_\text{2v}$ (red). The horizons $\pm k_\text{h}$ are marked by vertical lines.}}
	\label{figvel}
\end{figure}

The direction of travel of the modes is given by their group velocity, i.e.,
\begin{equation}
	v_g(k)=\frac{\partial \omega}{\partial k}=\frac{c \left(1+2\frac{k^2}{k_0^2}\right)}{\sqrt{1+\frac{k^2}{k_0^2}}},
\end{equation}
and considering the (negative) flow velocity $v_1$ or $v_2$ of each region, we obtain the following group velocities in the laboratory frame with respect to the flow $v_\text{lab}$ for counter-propagating (u) and co-propagating (v) modes
\begin{alignat}{3}
	\begin{aligned}
		v_\text{1u}(k)&=v_g(k)+v_1,& \quad v_\text{1v}(k)&=-v_g(k)+v_1, \\
		v_\text{2u}(k)&=v_g(k)+v_2,& \quad v_\text{2v}(k)&=-v_g(k)+v_2.
	\end{aligned}\label{vels}
\end{alignat}
Each of these velocities is shown in Fig. \ref{figvel}. As expected, velocities for co-propagating modes (v) are always negative (remember that flow velocity is negative). For counter-propagating modes (u) in the subsonic region ($v_1$), the velocity is always positive, but in the supersonic region ($v_2$) there is a finite region of $k$ where the velocity is negative. The region of opposite flow for $v_\text{2u}$ is marked by dashed lines in Fig. \ref{figvel} and the limiting values are known as horizons $\pm k_\text{h}$. For numerical calculations we will use dimensionless units. In this case, it means setting velocities $v$ in terms of the speed of sound, i.e., $v/c$.

\subsection{Horizon}
We define an horizon as a $k$-mode that is not moving with respect to the cavity, i.e., where the dispersion relation gives a group velocity $v_g(k)$ equal to the velocity of the external flow $v$. This is only possible for counter-propagating modes (u) in a supersonic region ($v_2$), i.e., only $v_\text{2u}(k)$ crosses zero, as seen in Fig. \ref{figvel}. Then, the horizon $k_\text{h}$ is given by
\begin{equation}
	\left.v_\text{2u}(k)\right|_{k=k_\text{h}}=0.
\end{equation}
This value defines the change of direction of counter-propagating waves in the supersonic region. The general solution is
\begin{equation}
	k_\text{h}=\pm\frac{k_0}{2^{3/2}}\sqrt{\frac{v^2}{c^2}-4\pm \frac{v}{c}\sqrt{\frac{v^2}{c^2}+8}}
\end{equation}
(see Eq. (24) in Ref. \cite{Larre2012}). For a sufficiently high-enough velocity, there are two real solutions, e.g., for velocity $v_2$ we obtain the values $k_\text{h}$ shown in Fig. \ref{figvel}, this is why $v_2$ is a supersonic velocity. For low velocities there are no real solutions, as in the case of $v_1$ in our example, this is correct, as there is no horizon for subsonic velocities.

\subsection{Transonic velocity}

As we just saw, for given $k_0$ and $\omega$ there is a minimum velocity needed to reach the horizon. This is the minimum velocity for which the solutions $k_\text{2ur}$ and $k_\text{2ul}$ exist (are real) and they are equal to each other, this is the ``transonic'' velocity $v_t$. This velocity is important, as it is the one that defines the subsonic ($v_t<v<0$) and supersonic regions ($v<v_t$); remember that $v$ and $v_t$ are negative.

This velocity can be found analytically, using an auxiliary function $q(k_0,\omega)$,
\begin{equation}
	q^3(k_0,\omega)=729k_0^4+270k_0^2\omega^2-2\omega^4+3^{3/2}|k_0|(27k_0^2-4\omega^2)^{3/2},
\end{equation}
as
\begin{equation}
	v_t^2=1+\frac{(2\omega)^{2/3}q(k_0,\omega)-2w^2}{6k_0^2}+\frac{(54k_0^2+\omega^2)|\omega|^{4/3}}{k_0^2 q(k_0,\omega)},
\end{equation}
and taking the negative square-root. For example, for $k_0=2$ and $\omega=1$, we obtain $v_t=-1.9002$.

\subsection{Norm}
Now, an important part of the Hawking effect relies on the mixing of positive and negative norms, for the following reason. Modes with positive norm carry annihilation operators, whereas modes with negative norm carry creation operators. A mixing of positive and negative norm modes thus corresponds to a mixing of annihilation and creation operators, which describes the creation of particles. The norm is given in terms of the scalar product \cite{Robertson2011}:
\begin{equation}
	(\phi_1,\phi_2) = -\ii\int_{-\infty}^{+\infty}\left( \phi_1^*(\partial_t + v\partial_x)\phi_2 - \phi_2(\partial_t + v\partial_x)\phi_1^*\right) \dd x \,.
	\label{scalar}
\end{equation}
The scalar product is a conserved quantity that corresponds to the particle number associated with a wave. It fulfills the following identities
\begin{equation}
	(\phi_1,\phi_2)=(\phi_2,\phi_1)^*,\qquad (\phi_1^*,\phi_2^*)=-(\phi_1,\phi_2)^*.
\end{equation}
Let us study the norm of our system for a region of constant $v$ \cite{Robertson2011}, also called flat profile configuration \cite{Larre2012}. We can write a general mode as a sum of plane waves $\exp(ikx-i\omega t)$ for  counter-propagating modes and $\exp(-ikx-i\omega t)$ for co-propagating ones. We obtain from the wave equation $\omega=\pm vk + F(k)$, where the $+$ corresponds to counter-propagating modes and the $-$ to co-propagating modes. Let us normalize the modes with respect to the scalar product, Eq.~(\ref{scalar}):
\begin{equation}
	\left(\ee^{\pm\ii k_1x-\ii (vk_1+F(k_1))t},\ee^{\pm\ii k_2x-\ii (vk_2+F(k_2))t}\right)=4\pi F(k_1)\delta(k_1-k_2).
\end{equation}
The norm is positive if $F$ is positive, which is the case for positive frequencies $\omega$ and co-propagating modes. For counter-propagating waves the norm is positive for positive frequencies $\omega$ if the phase velocity does not exceed the flow velocity $v$. Otherwise it is positive for negative $\omega$. For normalizing with respect to $\omega$ we express $\delta(k_1-k_2)$ as $v_\text{lab}\delta(\omega_1-\omega_2)$ with $v_\text{lab}$ from Eq. \eqref{vels}, and get
\begin{equation}
	\phi(x,t)=\frac{1}{\sqrt{4\pi|v_\text{lab}(k) F(k)|}}\ee^{\ii kx-i(vk+F(k))t}.
\end{equation}
In our case, we focus on the waves trapped between the horizons. We give those waves, the ur and ul waves, a negative norm, while u and v have positive norm. This means that our waves oscillate with positive frequencies $\omega$. In Fig. \ref{figbhl}(b), positive-norm modes are shown in solid arrows and negative-norm modes in dotted ones.

\section{Analytical solutions}
\begin{figure}
	\centering
	\includegraphics[scale=0.4]{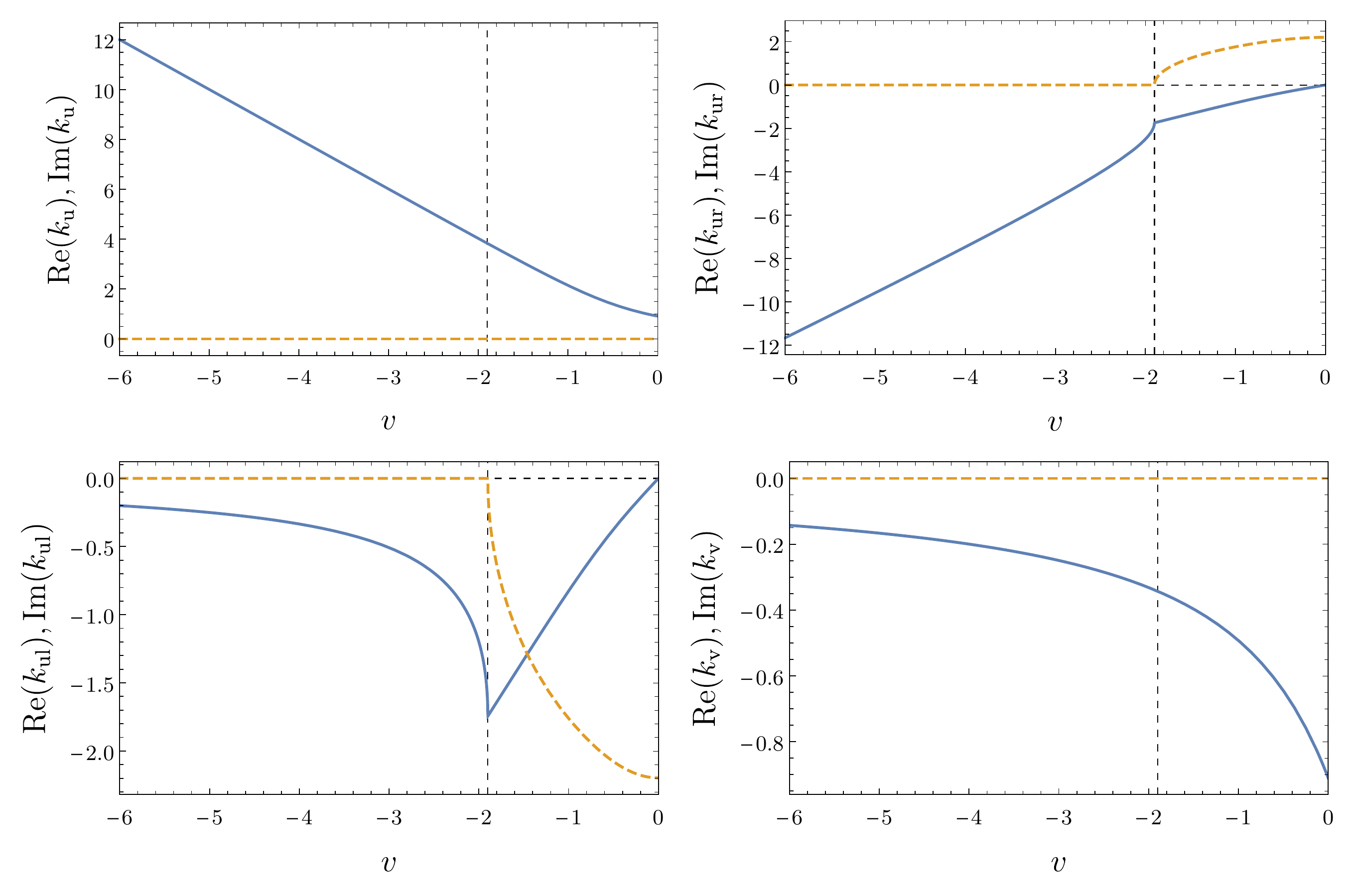}
	\caption{\small{The four analytic solutions changing the velocity $v$, its real (solid blue) and imaginary (orange dashed) parts for $k_0=2$ and $\omega=1$. The vertical lines correspond to the transonic velocity $v_t=-1.9002$. We can see that $k_\text{u}$ and $k_\text{v}$ are real for all velocities, while $k_\text{ur}$ and $k_\text{ul}$ are only for supersonic velocities.}}
	\label{figpgrid}
\end{figure}

\begin{figure}
	\centering
	\includegraphics[scale=0.45]{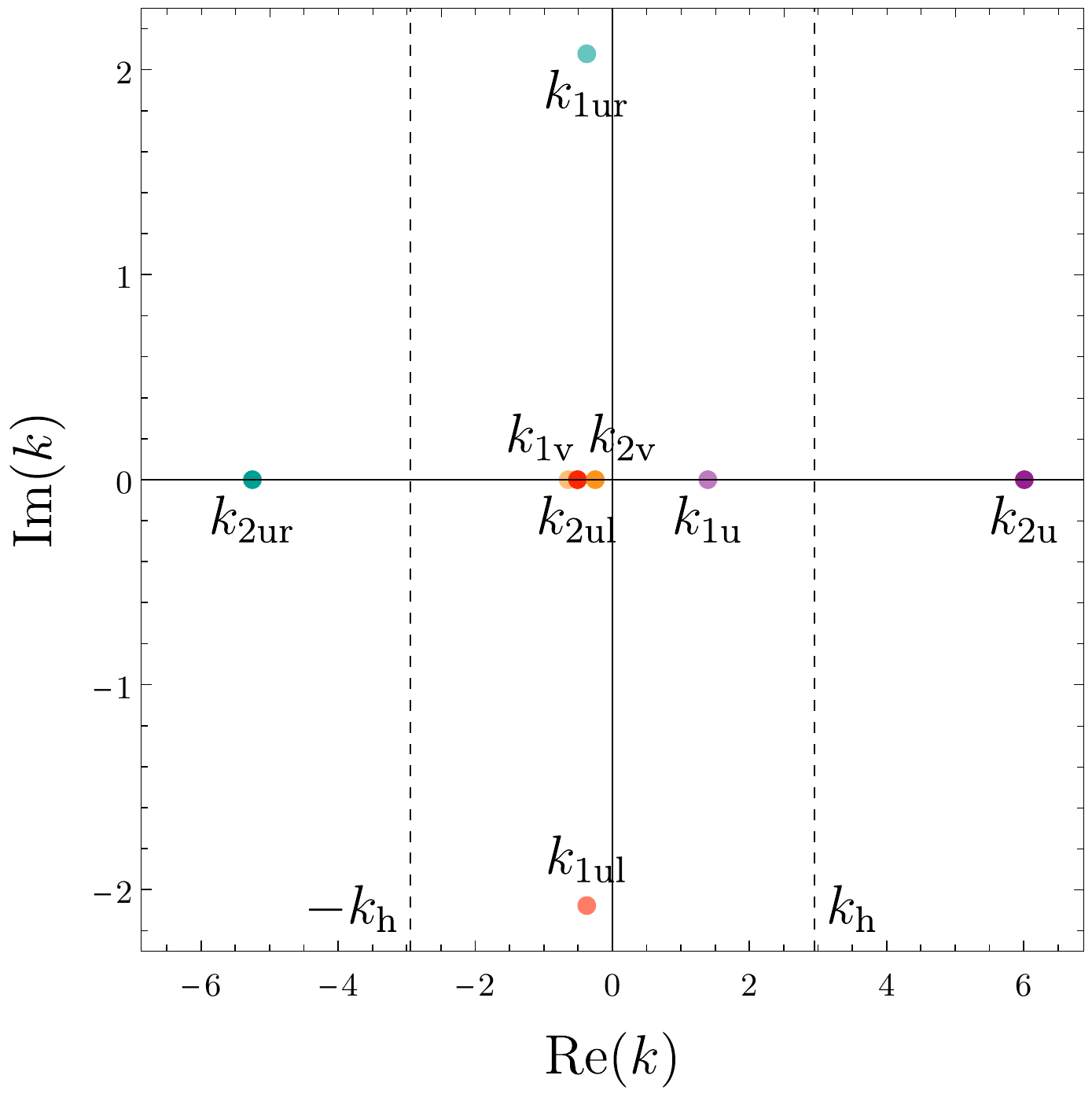}
	\caption{\small{Analytical solutions for $v_1=-1/2$ and $v_2=-3$ in the complex $k$-space. We obtain the original six real solutions plus two new complex ones. Lines at $k_\text{h}$ help us find the direction of travel for $k_\text{2u}$, $k_\text{2ur}$, and $k_\text{2ul}$.}}
	\label{figkspace0}
\end{figure}

We are interested in applying the theory of instabilities to this system. For that, we need to generalize the system to be valid for complex frequencies $\omega$ and wave numbers $k$. Therefore, we need to solve analytically the dispersion relation in Eq. \eqref{disp}. This is a quartic equation that can be written in canonical form
\begin{equation}
	k^4+d k^3+e k^2+f k +g=0,
\end{equation}
with the following coefficients:
\begin{equation}
	d=0, \quad e=\left(1-\frac{v^2}{c^2}\right)k_0^2,\quad f=\frac{2w_0vk_0}{c^2}, \quad g=-\frac{w_0^2k_0^2}{c^2}.
\end{equation}
This type of equations with $d=0$ can be reduced to an auxiliary cubic equation
\begin{equation}
	k^3 + \frac{e}{2} k^2 + \frac{e^2 - 4 g}{16} k - \frac{f^2}{64} = 0,
\end{equation}
and from its solutions $p_1,p_2,p_3$, one can obtain the solutions of the quartic equation we are interested in as
\begin{alignat}{3}\begin{aligned}
		k_1 &=\sqrt{p_1} + \sqrt{p_2} + \sqrt{p_3}, &\quad k_3 &= -\sqrt{p_1} + \sqrt{p_2} - \sqrt{p_3},\\
		k_2 &= \sqrt{p_1} - \sqrt{p_2} - \sqrt{p_3}, &\quad k_4 &= -\sqrt{p_1} - \sqrt{p_2} + \sqrt{p_3}.
\end{aligned}\end{alignat}
These are the analytical solutions of the dispersion relation \eqref{disp}, although the explicit solutions are too long to be written here, they can be used to perform analytical calculations. In this way we obtain four solutions for each part of the velocity flow, i.e., the supersonic and the subsonic ones. Plots of the four complex solutions for varying velocities are shown in Fig. \ref{figpgrid}. There we also marked the transonic velocity $v_t$, where solutions $k_\text{ur}$ and $k_\text{ul}$ become complex. The $k$ solutions for the subsonic $v_1=-1/2$ and supersonic $v_2=-3$ cases in Figs. \ref{figdisp} and \ref{figpgrid} are shown in the complex $k$ space in Fig. \ref{figkspace0}.

Comparing with the solutions of the numerical method, we obtain the same four solutions in the supersonic region and the two real solutions in the subsonic region. Furthermore, we obtain two more complex solutions that are complex conjugated of each other and that they do not appear in the usual treatment of the BHL system.

These two extra solutions in the subsonic part have complex $k$ and they will be exponentially growing to one side of $x$ and decaying to the other
\begin{equation}
	\ee^{\ii kx}=\ee^{\ii (k_R+\ii k_I)x}=\ee^{-k_I x}\ee^{\ii k_R x}.
\end{equation}
That is, if $k_I>0$ the solution will decay to the right, and if $k_I<0$ it will decay to the left.

\section{Black hole laser with complex frequencies}
\subsection{Qualitative and quantitative descriptions}
Now, we consider the condensate in a black-hole laser configuration as a quantum fluid. We would like to find its instabilities, i.e. the eigenfunctions of the system with complex frequencies $\omega=\omega_R+\ii\omega_I$ with positive imaginary part $\omega_I>0$, such that the amplitude will increase exponentially with time
\begin{equation}
	\ee^{-\ii \omega t}=\ee^{-\ii(\omega_R+\ii \omega_I)t}=\ee^{\omega_I t}\ee^{-\ii \omega_R t}.
\end{equation}

For this we take the BHL cavity as a whole and consider a situation where the only modes allowed to leave the cavity are the exponentially decaying ones, all these modes are in the subsonic region. From our labeling of the subsonic solutions, we can see that $k_\text{1ur}$ decays exponentially to the right and $k_\text{1ul}$ to the left.

\begin{figure}
	\centering
	\includegraphics[scale=0.3]{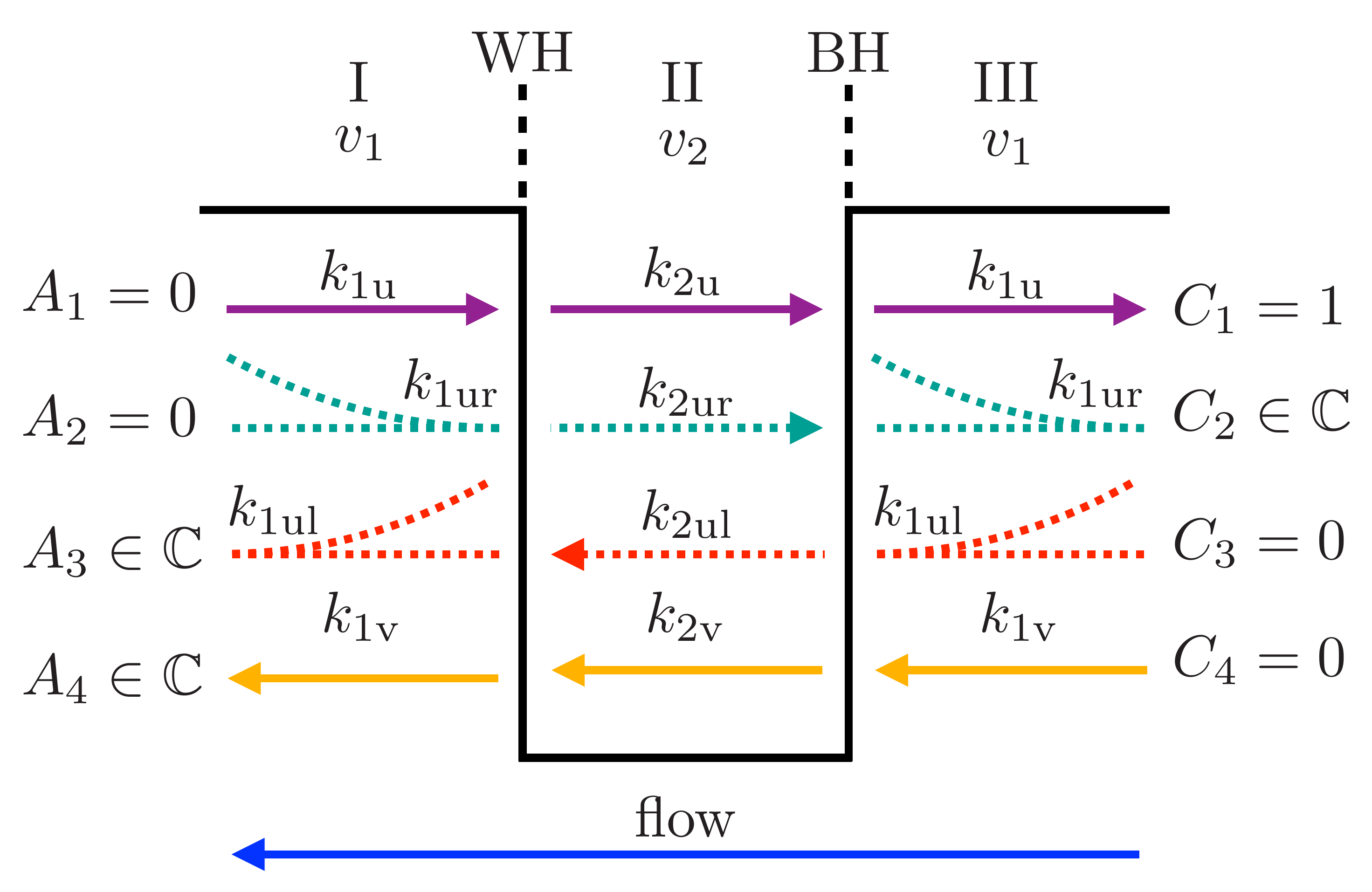}
	\caption{\small{Diagram of modes of a BHL with instabilities. Modes $k_\text{1ur}$ and $k_\text{1ul}$ are exponentially decaying or growing to one side. The shown coefficients $A_j$ and $C_j$ describe the situation with square-integrable solutions.}}
	\label{figbhl2}
\end{figure}

We expect that adding a small positive imaginary part to the frequency will not change the qualitative behavior of our solutions from real frequencies to keep having a black-hole laser configuration. Taking into account the previous considerations, we have a new configuration for the BHL with complex frequencies in Fig. \ref{figbhl2} that includes new complex solutions.

Let us state this situation in equations. We divide the quantum field $\phi$ into the three regions I, II, and III, as shown in Fig. \ref{figbhl2}. In each region, the wave function of the fluid $\phi(x)$ is given by
\begin{equation}
	\phi_1=\mathbf{A}\cdot \ee^{\ii\mathbf{k}_1 x},\quad
	\phi_2=\mathbf{B}\cdot \ee^{\ii\mathbf{k}_2 x},\quad
	\phi_3=\mathbf{C}\cdot \ee^{\ii\mathbf{k}_1 x},
\end{equation}
where
\begin{equation}
	\mathbf{A}=(A_1,A_2,A_3,A_4),\quad
	\mathbf{B}=(B_1,B_2,B_3,B_4),\quad
	\mathbf{C}=(C_1,C_2,C_3,C_4),
\end{equation}
and
\begin{equation}
	\ee^{\ii \mathbf{k}_1\cdot\mathbf{x}}=
	(\ee^{\ii k_{11} x},\ee^{\ii k_{12} x},\ee^{\ii k_{13} x},\ee^{\ii k_{14} x}),\quad
	\ee^{\ii \mathbf{k}_2\cdot\mathbf{x}}=
	(\ee^{\ii k_{21} x},\ee^{\ii k_{22} x},\ee^{\ii k_{23} x},\ee^{\ii k_{24} x}).
\end{equation}
In the laser configuration, incoming and outgoing modes are allowed as long as they are exponentially decaying away from the cavity. This means that the BHL configuration is $A_1=A_2=C_3=C_4=0$, as stated in Fig. \ref{figbhl2}. The other coefficients must be found to define a BHL mode.

\subsection{Transfer matrix method}

In order to solve this problem we use the transfer matrix method. We can use this method because in our system, the velocity profile is constant in all regions except on the interfaces, where we should ensure the continuity of $\phi(x),\phi'(x),\phi''(x),\phi'''(x)$. This is because we have a quartic equation. We can define the transfer matrices from region I to II ($M_1$) and II to III ($M_2$) from the matrices with coefficients from $\phi(x),\phi'(x),\phi''(x),\phi'''(x)$:
\begin{equation}
	m_1=\begin{pmatrix}
		1 & 1 & 1 & 1 \\
		\phantom{-}\ii k_{11} & \phantom{-}\ii k_{12} & \phantom{-}\ii k_{13} & \phantom{-}\ii k_{14} \\
		-k_{11}^2 & -k_{12}^2 & -k_{13}^2 & -k_{14}^2 \\
		-\ii k_{11}^3 & -\ii k_{12}^3 & -\ii k_{13}^3 & -\ii k_{14}^3
	\end{pmatrix},\quad
	m_2=\begin{pmatrix}
		1 & 1 & 1 & 1 \\
		\phantom{-}\ii k_{21} & \phantom{-}\ii k_{22} & \phantom{-}\ii k_{23} & \phantom{-}\ii k_{24} \\
		-k_{21}^2 & -k_{22}^2 & -k_{23}^2 & -k_{24}^2 \\
		-\ii k_{21}^3 & -\ii k_{22}^3 & -\ii k_{23}^3 & -\ii k_{24}^3
	\end{pmatrix}.
\end{equation}
The first subindex corresponds to the velocity $v_1$ or $v_2$. The second one specifies the solution, from 1 to 4. We can choose any order of our $k$ solutions as long as we are consistent; we choose the ordering u, ur, ul, and v, as shown in Fig. \ref{figbhl2}.

Then, the transfer matrices are simply
\begin{equation}
	M_1=m_2^{-1}m_1, \quad M_2=m_1^{-1}m_2=M_1^{-1}.
\end{equation}
We also define the propagation matrices to take the solution between the interfaces of the cavity. They are defined by
\begin{align}
	P_L&=\text{diag}(\ee^{-\ii k_{21}L},\ee^{-\ii k_{22}L},\ee^{-\ii k_{23}L},\ee^{-\ii k_{24}L}),\\
	P_R&=\text{diag}(\ee^{\ii k_{21}L},\ee^{\ii k_{22}L},\ee^{\ii k_{23}L},\ee^{\ii k_{24}L}).
\end{align}

Finally, the total transfer matrix ($M$) from region III to I is given by
\begin{equation}
	M=M_1 P_L M_2= m_2^{-1}m_1 P_L m_1^{-1}m_2,
\end{equation}
similarly, from region I to III we need to use $M^{-1}$.

Using $M$ we can obtain the coefficients $A_1,A_2,A_3,A_4$ in terms of $C_1,C_2,C_3,C_4$ for a given complex frequency $\omega$. Each complex $\omega$ has a set of four complex wave-number solutions $k$ for the subsonic region $\mathbf{k}_1$ and another four for the supersonic region $\mathbf{k}_2$. Together they all determine the $A_j$ coefficients.
\begin{equation}
	\phi_\text{I}(x)=M\phi_{\text{III}}(x).
\end{equation}
This problem can be solved analytically. Starting from the four coefficients $\mathbf{A}$, for any complex frequency $\omega$ one can calculate the corresponding eight complex wave-numbers $\mathbf{k}_1(\omega)$ and $\mathbf{k}_2(\omega)$, and finally calculate the four coefficients $\mathbf{C}$.

\subsection{Spontaneous lasing modes}
\begin{figure}
	\centering
	\includegraphics[scale=0.5]{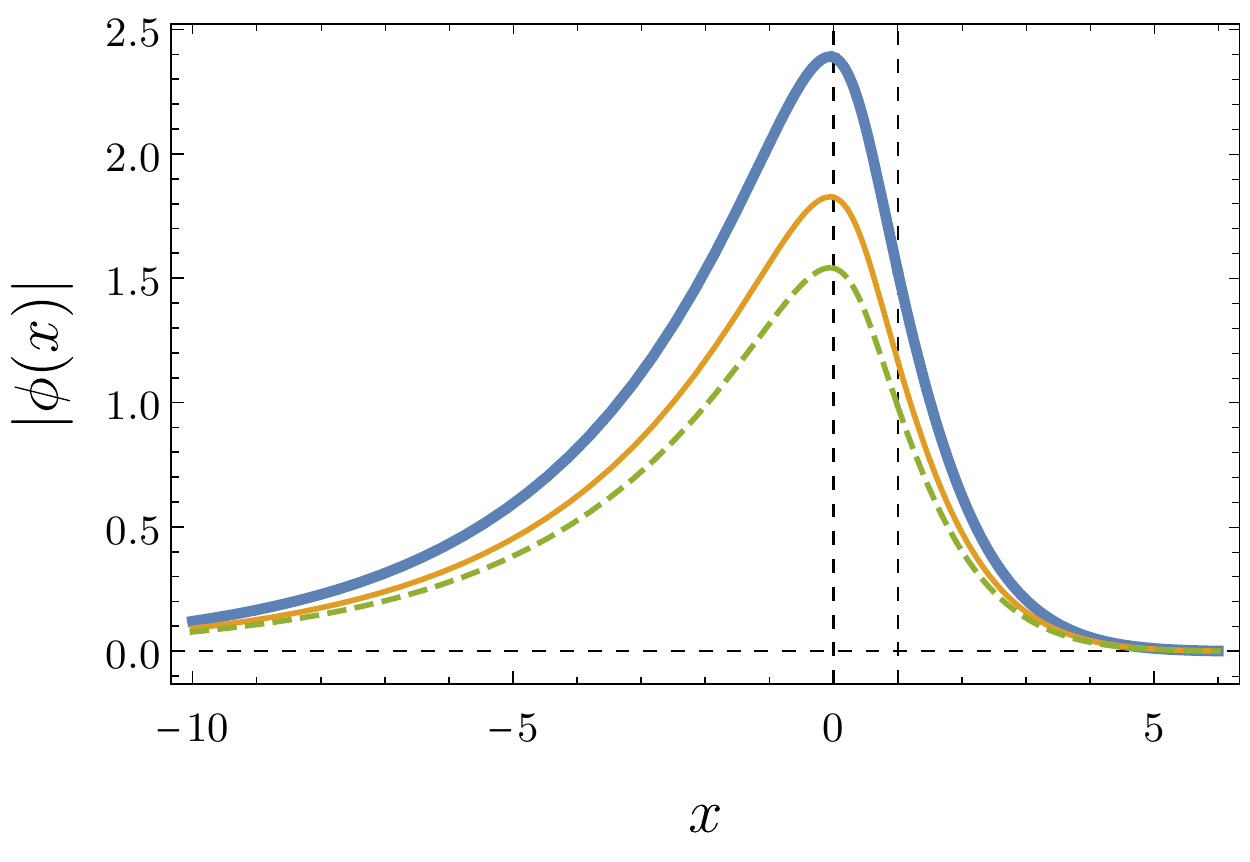}\hspace{5mm}
	\includegraphics[scale=0.5]{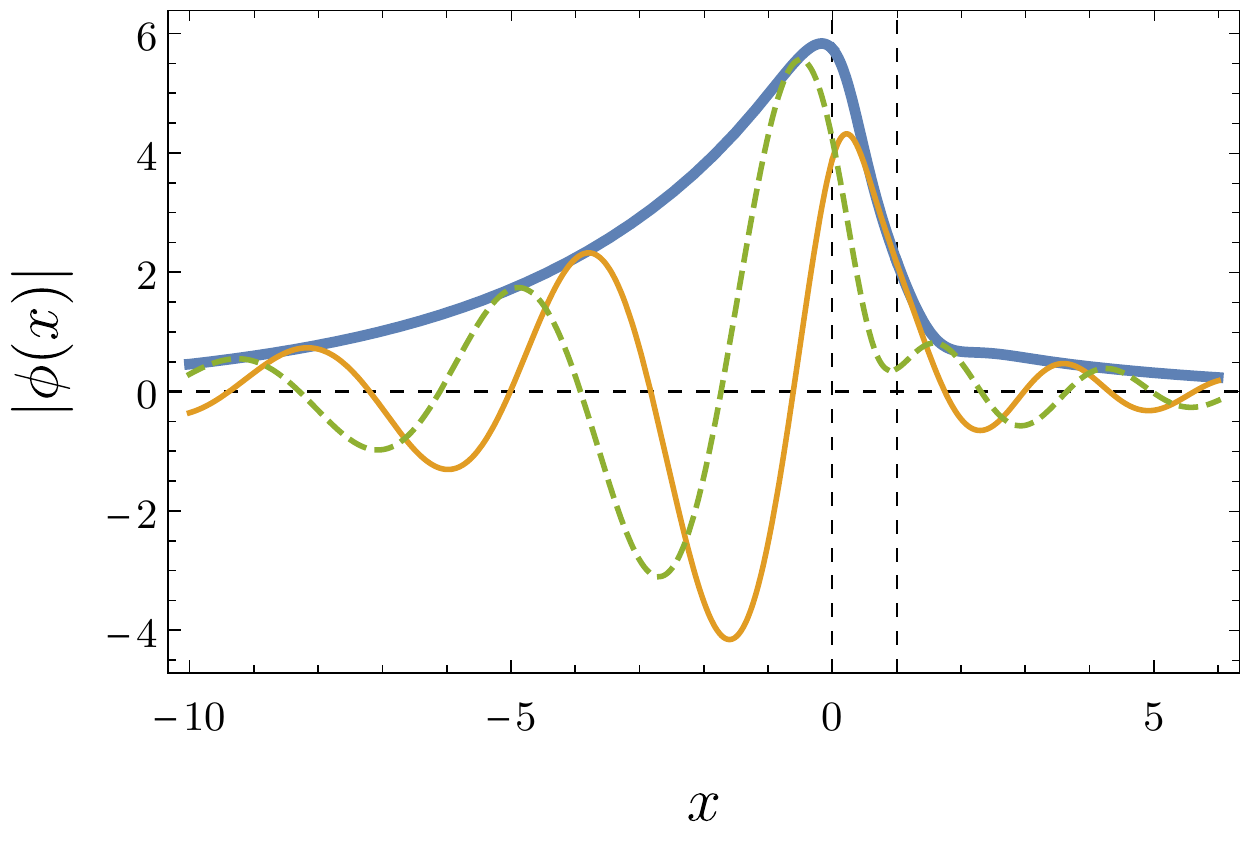}
	\caption{\small{Quantum field $\phi(x)$ for a cavity formed between $x=0$ and $x=L=1$. We plot $|\phi|$ (thick solid blue), Re$(\phi)$ (thin solid orange), Im$(\phi)$ (dashed green) for $v_2=-2$ (left) and $v_2=-3$ (right).}}
	\label{figcont}
\end{figure}

We define a spontaneous lasing mode as a mode of the BHL with complex eigenvalues that has a square-integrable stable solution. This is the situation shown in Fig. \ref{figbhl2}, where $A_1=A_2=C_3=C_4=0$. We start with $\phi_\text{III}(x)$ given by the coefficients $\mathbf{C}=(1,\rho,0,0)$ with $\rho\in\mathbb{C}$ to be found (we also fix its phase). Then, we obtain $\phi_\text{I}(x)$ using the transfer matrix method such that $A_1=A_2=0$. The solution with one null coefficient can be found analytically, but the second one has to be solved numerically. For example, for $v_1=-1/2$ and $v_2=-3$, there is a single solution given by $\omega_\ell=2.4613 + 0.5667 \ii$, where the subindex $\ell$ indicates it is a lasing mode. The quantum fields $\phi(x)$ for this and an additional solution are shown for the three regions in Fig. \ref{figcont}, where we see that their absolute value and real and imaginary part are continuous and decaying away from the cavity. We also checked that the first three derivative are continuous at $x=0$ and $x=L$. 

\begin{figure}
	\centering
	\includegraphics[scale=0.45]{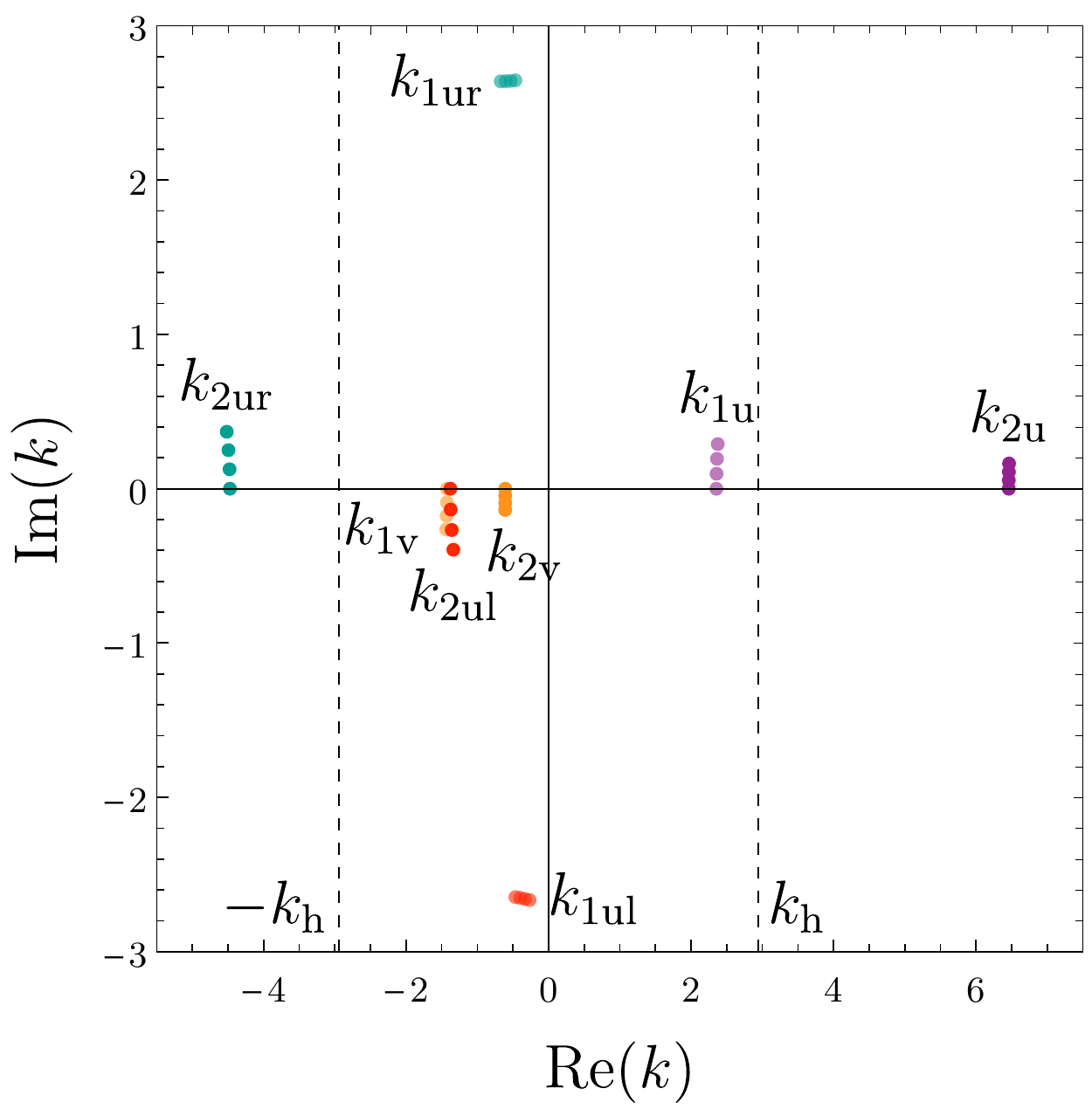}
	\caption{\small{Change of $\mathbf{k}_1$ and $\mathbf{k}_2$ for fixed Re$(\omega_\ell)$ and increasing value of Im$(\omega)$ from 0 to Im$(\omega_\ell)$ in fixed steps for $v_1=-1/2$ and $v_2=-3$.}}
	\label{figcomplex}
\end{figure}

We need to check that the qualitative behavior of our new complex solution follows our stated problem. In particular, we need to check that modes $k_\text{2ur}$ and $k_\text{2ul}$ decay on the correct side of the cavity, as the ordering of those modes is initially arbitrary. The eights values of $\mathbf{k}_1$ and $\mathbf{k}_2$ are shown in Fig. \ref{figcomplex} with fixed real part Re$(\omega_\ell)$ and increasing imaginary part from 0 to Im$(\omega_\ell)$ in fixed steps.

As we have pointed out in the introduction, the exponential decay of the spatial modes does not mean that the radiation is trapped inside the BHL. On the contrary, it is propagating out of the cavity. Figure \ref{figcont} shows that  the spatial shape is asymmetric: most of the radiation moves out as co-propagating $v$-waves in region I. Reflection is a common feature in wave propagation across sharp boundaries, as in our case. So the large co-propagating component is not surprising. The surprise is the large trans-Planckian contribution to the emitted radiation, usually larger than the usual counter-propagating component. For example, for $v_1=-1/2$ and $v_2=-3$, the coefficients outside the cavity for the single lasing mode are $\mathbf{A}=(0,0,-1.2149 +\ii 0.3174, 5.1095 + \ii 3.9008)$ in region I, $\mathbf{C}=(1,1.1497 + \ii 0.3824,0,0)$ in region III. This shows that trans-Planckian radiation is able to tunnel out of the BHL at a comparable rate as the ordinary Hawking radiation. 

Numerically, a single solution can be found for the slowest supersonic velocities $v_2$, i.e., close to $v_t$, e.g., $v_2=-3$. However, we can find more solutions if we choose an even more negative value, e.g., for $v_2=-5$ there are two solutions. Is there another way of explaining these findings? In the next section we will compare our solution found through the theory of instabilities with a simple model.

\section{Comparison with simple model}
In previous sections we show how to find spontaneous lasing modes of the BHL using the theory of excitations. Varying $v_1$ and $v_2$, we found numerical solutions with one or more lasing modes. We now use a simple mode to get a better picture of the dynamics of the lasing mode. This model considers plane-waves solutions, i.e., we take the solutions for our lasing mode $k(\omega_\ell)$ with complex frequency $\omega_\ell=\omega_R+\ii \omega_I$, but we consider that the imaginary part is zero $\omega_I\rightarrow 0$. This is a plane-wave model with real wave numbers $k(\omega_R)$.

According to the usual BHL evolution (see Fig. \ref{figbhl}(a)), there are two trapped modes inside the cavity, one traveling to the left $k_\text{2ul}$ and one to the right $k_\text{2ul}$. These modes stimulate a mode with opposite norm that is able to leave the cavity $k_\text{1u}$ and at the same time amplifying the next cycle of trapped modes. With the plane-wave model, we check if the wave-number solutions $k$ follow a round-trip condition for a resonance.

\subsection{Resonance condition}

A resonance occurs when the phase difference of the trapped radiation after each round-trip is an integer multiple of $2\pi$. As two reflections at turning points occur during each round-trip, the phase mismatch  between $k_\text{2ul}$ and $k_\text{2ur}$ should be $\Delta k=(2n+1)\pi$.

\begin{figure}
	\centering
	\includegraphics[scale=0.55]{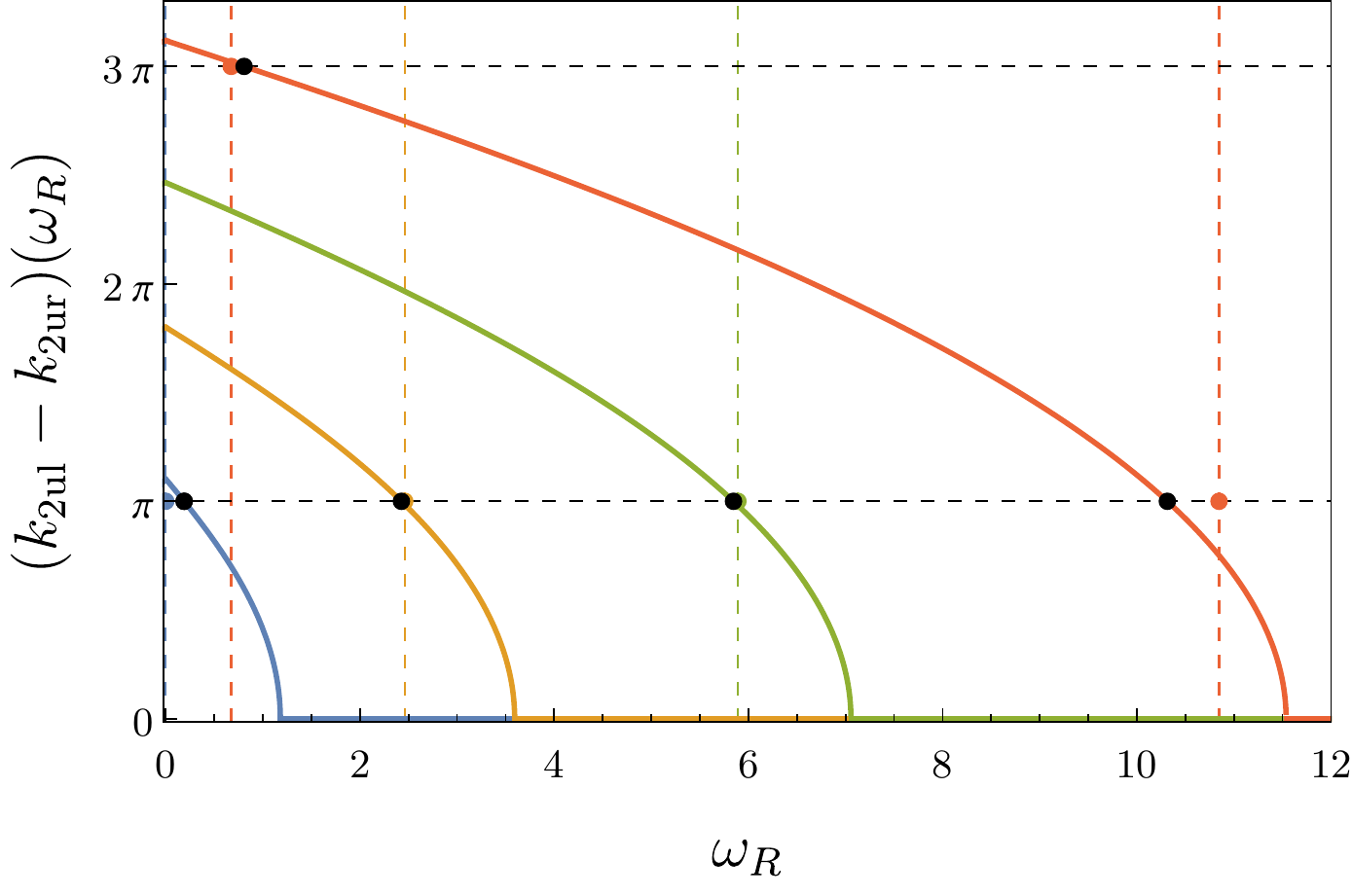}
	\caption{\small{Phase difference $(k_\text{2ul}-k_\text{2ur})(\omega_R)$ considering the plane-wave model with $v_2=$ $-2$ (blue), $-3$ (orange), $-4$ (green), and $-5$ (red). The phase difference of $\pi$ and $3\pi$ predicted by the model are marked by black points. The dashed lines mark the $\omega_R$-value found by the theory of excitations in the corresponding color and a point marks its intersection with the phase value.}}
	\label{figphase}
\end{figure}

In Fig. \ref{figphase} we show the phase difference $(k_\text{2ul}-k_\text{2ur})$ at $\omega_\ell$ using the plane-wave model ($\omega_I=0$) for different values of $v_2$. We show in black the predictions of the plane-wave model and, for comparison, we show as vertical dashed lines the solutions of $\omega_\ell$ found with the theory of excitations. All of the predictions of the simple model are close to the actual solution of the full system and, more importantly, the number of roots is given by this model.

This means that the plane-wave model for real frequencies is good enough to give us an approximation for $\omega_R$. Moreover, it enforces our interpretation of the complex solution obtained using the theory of excitations. Also, it agrees with the result of the last equation of Ref. \cite{Leonhardt2007} using the plane-wave model. Indeed, we are finding that the continuous bouncing of modes $k_\text{2ul}$ and $k_\text{2ur}$ is responsible for the creation of particles by the Hawking-like process through instabilities.

%{\color{red} \subsection{Amplification rate}
%Using the theory of excitations we obtain more information of our process, i.e., the rate of particle creation that is proportional to $\omega_I$. This value is not given by the phase-matching condition. However, it can be obtained by considering two factors: (1) the time that it takes a trapped signal to complete a period $k_\text{2ul}\rightarrow k_\text{2ur}\rightarrow k_\text{2ul}$ and (2) the change of norm after one period.

%The period can be easily calculated from the cavity length ($L$) and the velocity of the waves (see Eq. \eqref{vels} and Fig. \ref{figvel}). The velocity for the involved modes ($k_\text{2ul}$ and $k_\text{2ur}$) is given by $v_\text{2u}(k)$. Then, the period ($T$) is given by
%\begin{equation}
%	T=\frac{L}{|v_\text{2u}[k_\text{2ul}(\omega_R)]|}+\frac{L}{|v_\text{2u}[k_\text{2ur}(\omega_R)]|}.
%\end{equation}
%}

\section{Conclusions}
In this paper, we applied the theory of excitations to describe a black-hole laser in a Bose-Einstein condensate. By using this theory, we are able to find spontaneous lasing modes, where a system without incoming signal is able to exponentially amplify the trapped modes. This is a signature of a Hawking-like process that occurs here as eigenmodes with complex frequencies.

Moreover, we found that the black-hole laser is well understood \cite{Corley1999,Leonhardt2007} using a simple plane-wave model for real frequencies. We can take the real part of our frequency solutions $\omega_\ell$ and use this simple model to compare with the result obtained from the theory of excitations. From this we obtain that the resonance condition for the appropriated modes gives a good approximation of Re$(\omega_{\ell})$. %Secondly, the amplification rate can be obtained with the simple model by calculating the norm change of a trapped mode in a period, this approximates Im$(\omega_\ell)$ to a good degree.

Our theory is thus consistent with a simple, intuitive model based on waves with real frequencies. Yet it also shows a surprising feature for instabilities with complex frequencies: a substantial part of the emitted radiation consists of trans-Planckian waves. For real frequencies, these waves would not be able to propagate outside of the horizons, but for complex frequencies they would reach an observer in the same way as regular Hawking radiation. 

\section*{Acknowledgement}
DB acknowledges the financial support of Conacyt (Mexico) project A1-7751. UL was funded by the European Research Council and the Israel Science Foundation.

\bibliographystyle{ieeetr}
\bibliography{references}
\end{document}